\documentclass{aa}
\usepackage{graphicx,epsf,epsfig,float}
\usepackage{txfonts}
\usepackage{longtable}
\usepackage{lscape}
\usepackage{natbib}
\bibpunct{(}{)}{;}{a}{}{,}

\usepackage{color}

\def\nom{Gon{\c{c}}alves}
\def\Agata{R\' o\. za\' nska~}

\begin{document}
   \title{How to estimate the distance to the Warm Absorber in AGN \\ 
    from photoionized models}


   \author{A.~\Agata\inst{1}
          \and
          I. Kowalska\inst{2}
	  \and
	  A. C.~\nom\inst{3,4,5}
	  }

   \offprints{A.~\Agata}

   \institute{
Copernicus Astronomical Center, Bartycka 18, 
00-716 Warsaw, Poland 
(\email{agata@camk.edu.pl}) 
         \and
Warsaw University Observatory, Al. Ujazdowskie 4, 
00-478 Warsaw, Poland 
         \and
Observatoire Astronomique de Strabsourg, Universit\'e 
Louis Pasteur, CNRS, 11 rue de l'Universit\'e, 
67000 Strasbourg, France
        \and
LUTH, Observatoire de Paris, CNRS, 
Universit\'e Paris Diderot, 5 Place Jules Janssen, 
92190 Meudon, France  
         \and 
CAAUL, Observat\'orio Astron\'omico de Lisboa,
Tapada da Ajuda, 1349-018 Lisboa, Portugal 
         }
   \date{Received \today ; accepted (to be inserted later)}

   \abstract
{} 
{We present a method that allows us to estimate the distance from the
continuum source located in the center of AGN to the 
highly ionized gas called warm absorber.} 
{We compute a set of constant total pressure photoionization models
compatible with the warm absorber conditions,
where a metal-rich gas is irradiated
by a continuum  in the form of a double
power-law. The first power-law is hard up to 100 keV and represents
a radiation from an X-ray source, while the second power-law extends downwards from 
several eVs and illustrates a 
radiation from an accretion disk. }
{When the ionized continuum is dominated by the soft component, 
the warm absorber is 
 heated by free-free absorption, instead of Comptonization, and the
transmitted spectra show different absorption line characteristics for 
 different values of the hydrogen
number density at the cloud illuminated surface. }  
{This fact results in the possibility of deriving
the number density at the cloud illuminated side 
 from observations and hence the distance to the warm absorber.} 


\titlerunning{Distance to the Warm Absorber in AGN}
\authorrunning{A.~\Agata et al.}

   \maketitle


\section{Introduction}
Many Active Galactic Nuclei (AGN) exhibit 
numerous absorption features of highly ionized material
in their UV/X-ray spectra.
Such gas, called Warm Absorber (hereafter WA), is located on the line-of-sight
towards the observer and it is illuminated by radiation 
originating from the active nucleus.

Most of the observed absorption lines are blushifted, suggesting that the
metal-rich gas is outflowing.  
We still don't know how such a wind is powered and where it is 
lunched, 
nevertheless, due to high resolution spectroscopic 
observations of AGN from {\it FUSE}, {\it Chandra}, {\it XMM} 
and other satellites,
we are able to make some diagnostics on the physical conditions
of the WA \citep[for a review see:][]{kriss2004,blustin2005,rozanska2007}.

The observed absorption lines velocity shifts
are of the order of $10^4$ km s$^{-1}$ in case of UV absorbers 
\citep{gabel2003} and  of the order of $10^2$ to $10^3$ km s$^{-1}$ 
in the case of X-rays
\citep{kaspi2001,kaastra2002}.
The column density of the WA is generally estimated to be about 
$10^{21-23}$ cm$^{-2}$. The absorbing gas 
comprises different ionization phases, 
corresponding to temperatures from about
$10^5$ K, when iron is partially ionized, up to 
a few $10^7$ K, when iron is almost completely ionized 
\citep{netzer2003,krongold2003,steenbrugge2005}.
From the above hints, photoionization modeling 
of the WA is made, trying to answer to the most important and 
unsolved question: 
how far from the continuum source is the absorbing gas located? 

However, there is one  difficulty which prevents us from answering this 
question. It is well known that photoionized models of a cloud 
illuminated by a single X-ray power-law, typically ranging 
from 0.01 keV up to 100 keV (hereafter we will call this a hard
X-ray illumination)  are degenerate \citep{rozanska2007}. We
cannot distinguish between clouds with the same ionization parameter, 
but different hydrogen number density $n_0$, at the cloud illuminated surface, 
and at different locations.
The transmission spectrum from a rarefied cloud 
($n_0 \sim 10^5$ cm$^{-3}$) located farther away i.e. at $\sim 0.1$ pc, 
is identical to the  spectrum 
of a dense cloud ($n_0 \sim 10^{10}$ cm$^{-3}$), situated at $\sim 0.0001$
pc  from a continuum source.  
The first case is consistent with the WA being co-spatial with the 
NLR (narrow line
region) or dusty torus, while the second case corresponds to the 
closest neighbourhood
of an accretion disk and BLR (broad line region). 
 
Some estimations on the distance to the WA were done 
using variability studies  \citep{netzer2003,krongold2005}. 
However, using the same 900 ksec {\it Chandra} data of NGC 3783,
 \citet{netzer2003} found upper 
limits for the location of three ionization phases to be 3.2, 0.63, and 0,18
pc respectively, while \citet{krongold2005} estimated the location of 
the high ionization phase at 0.0029 pc and low ionization phase at 
0.0004-0008 pc. 

In this paper we show that the degeneracy of photoionized models
breaks down when the WA is illuminated by an AGN continuum
including the thermal disk radiation,
represented here as a second power-law component.
The disk contribution is
particularly important in the case of quasars, since their 
broad band spectra are clearly dominated by the soft disk emission.
In this article we explain why double power-law models are not
degenerate and claim that using such models provides a way
to determine the distance to the WA in some objects.

The structure of the paper is as follows:
In Sec.~\ref{sec:odl} we explain how to estimate the distance to the 
WA from photoionized models. 
Section~\ref{sec:mod} describes the photoionization models used in this
work. Results for different spectral shapes are presented in
Sec.~\ref{sec:res}. Discussion of the models is presented in
Sec.~\ref{sec:dis}, and main conclusions of our work are drown in  
Section~\ref{sec:con}.

\section{The Distance to the Warm Absorber}
\label{sec:odl}

The distance from the central engine to the WA is 
a clue issue for understanding the wind geometry. 
We can derive this distance from observations using photoionized
models, which are parametrized by the so called ionization parameter,
defined at the cloud surface. 
There are three ionization parameters being used by different groups and 
photoionization codes, but the 
idea is the same: to define the amount of ionized radiation
which reaches the cloud.
 Here, we discuss the problem of distance derivation 
using the ionization parameter defined as:
\begin{equation}
 \xiup = \frac{L_{ion}}{n R^2}, 
\label{eq:ion}
\end{equation} 
where $L_{ion}$ is the ionizing source luminosity, $n$ is
the hydrogen number density,
and  $R$ is the distance from  an illuminating source.
Nevertheless, the problem of distance derivation does not depend 
on the ionization parameter used by different
groups,  and refers to the value of this parameter on the cloud 
surface i.e. for intrinsic luminosity, 
and for surface hydrogen number density, hereafter $n_0$.

From observations, we are able to get the luminosity of 
the object as a separate measurement, 
and the ionization parameter on the cloud surface is obtained by fitting 
the data with a given photoionization model. 
But we are still left with the $n_0 R^2$ product, meaning the distance 
to the WA depends on the value of the number density at the cloud 
illuminated surface. 

Below, we show how to determine the number density at the illuminated side
of the WA using photoionization modeling.
Then, after simple inversion of Eq.~\ref{eq:ion}, 
we obtain the distance to the absorber.   
 
\section{Photoionization modeling}
\label{sec:mod}

In this paper all photoionization models are computed using the {\sc
titan} code developed by \citet{dumont2000,collin2004}. 
{\sc titan} is well-suited both for the study of optically thick
and thin media, such as the WA. 
It computes the gas structure in thermal and
ionization equilibrium assuming non-LTE, and it provides the
reflected, emitted
outward, and absorbed spectra. {\sc titan} treats the transfer of both 
the lines and the continuum using the ALI (Accelerated Lambda Iteration)
method, which precisely computes line and continuum fluxes in a
self-consistent way. 
Our atomic data include $\sim 10^3$ lines from ions and atoms 
of H, He, C, N, O, Ne, Mg, Si, S, and Fe, all elements having cosmic 
abundances. 

The total (gas+radiation) pressure is computed after the plasma temperature
has been determined by the thermal balance equation. The pressure is
kept constant within the medium thus allowing to determine the density profile.
The model is iterated until convergence. 
We have shown that the assumption of constant pressure allows the
ionized gas to be naturally stratified due to illumination  
\citep{rozanska2006,goncalves2006}; as a result, the 
cloud  comprises different ionization stages which is 
consistent with observations.   

In all our models we assume  a plane-parallel geometry, with the WA medium 
being illuminated on one side by a given incident
continuum.  We assume the external radiation to hit the cloud 
perpendicularly to its surface. 

In the frame of photoionization modeling of the warm absorber,  
there is no possibility to reconstruct velocity structure observed 
in some cases \citep[e.g.][]{steenbrugge2005},where each
ion has slightly different velocity. We can only 
consider the case where all ions have the same turbulent velocity,
and it is taken into account in our code.
 
The models are parametrized by the ionization parameter $\xiup$ at the
cloud surface, the total column density $N_H$, the hydrogen number density
at the cloud illuminated surface 
$n_0$, and the incident flux Spectral Energy Distribution (SED).

The parameter $\xiup$ is very well constrained on the 
surface of an illuminated medium. But if we want to trace ionization 
properties deeper in the gas, we use the so called dynamical ionization parameter
defined as:
\begin{equation}
\Xi =  \frac{\xiup}{4\pi c k T} = \frac {L_{ion}}{4\pi c R^2}
\frac{1}{n k T}= \frac {F_{ion}}{c P_{gas}} = \frac {P_{rad}}{P_{gas}},
\label{eq:ion}
\end{equation} 
where $c$ is a velocity of light, $T$ is 
temperature of the gas, $F_{ion}$ is a flux which affects the
cloud, and $P_{rad}$ and $P_{gas}$ are radiation and gas pressure 
respectively. Below, we describe the ionization structure of presented clouds 
using relation $\Xi(T)$.
  
In this paper we consider two possible shapes of SED first: a single power-law 
roughly representing the spectra of Seyfert galaxies,
and second: a double power-law representing the spectra of disk dominated 
bright quasars. In both cases,
the power-laws are modified by exponential high- and low-energy
cut-offs according to the formula:
\begin{equation}
F_{E} = E^{-\alpha}exp(-E/E_{max})exp(-E_{min}/E),
\end{equation}
where $F$ is the flux, $E$ - energy, $\alpha$ - spectral index, 
$E_{min}$ and $E_{max}$ are low and high spectral ranges
respectively.
In the case of double power-law models, we have additional 
parameter, which is the relative
normalization of both power-law components, i.e. the ratio of 
their integrated fluxes denoted as $F_X/F_{soft}$. 
This relative normalization and exponential cut-offs 
ensure that there is no discontinuity at the point, where
two power-laws merge.

\section{Results for the different shape of illuminated continuum}
\label{sec:res}

To illustrate the determination of distance by photoionized 
models, we consider three clouds of different hydrogen density 
number at the illuminated surface: 
$n_0=10^6$,  $10^8$, and  $10^{10}$ cm$^{-3}$.  
The ionization parameter on the cloud surface is the same  in  
the three cases and 
$\xiup=10^5$. The total column density of each cloud is taken to be
close to the maximum column density for which the temperature profile 
is thermally stable for a given spectral energy distribution
\citep[see][]{rozanska2007}. In all cases, the  
total column density is of the order of $10^{23}$ cm$^{-2}$,  
 and we assume the turbulent velocity of the gas equaled
300 km/s. 

\begin{figure*}
\epsfxsize=6.1cm \epsfbox[20  555 210 690]{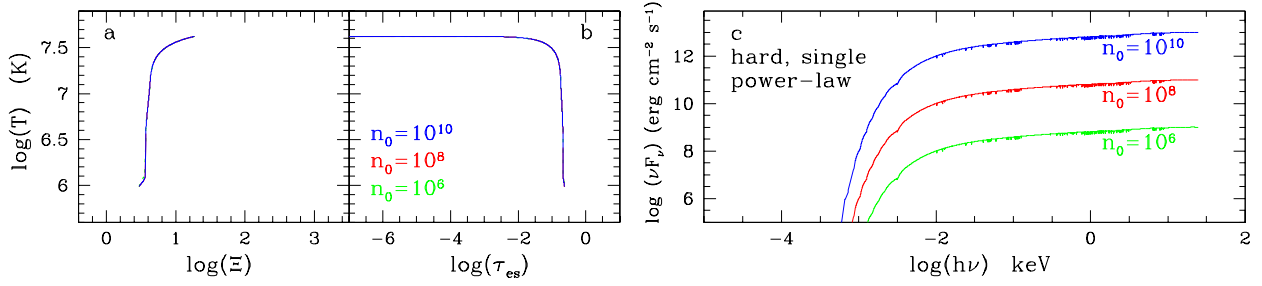}

\caption{ Ionization structure (a), 
temperature structure (b),  and transmitted spectrum (c) for
a WA illuminated by a hard single power-law continuum for
three different values of $n_0$: 
$10^6$~cm$^{-3}$ (green line), $10^8$~cm$^{-3}$ (red line), and  
$10^{10}$~cm$^{-3}$ (blue line).}
\label{fig:sing}
\end{figure*}

\begin{figure*}[t]
\epsfxsize=6.1cm \epsfbox[20 380 210 690]{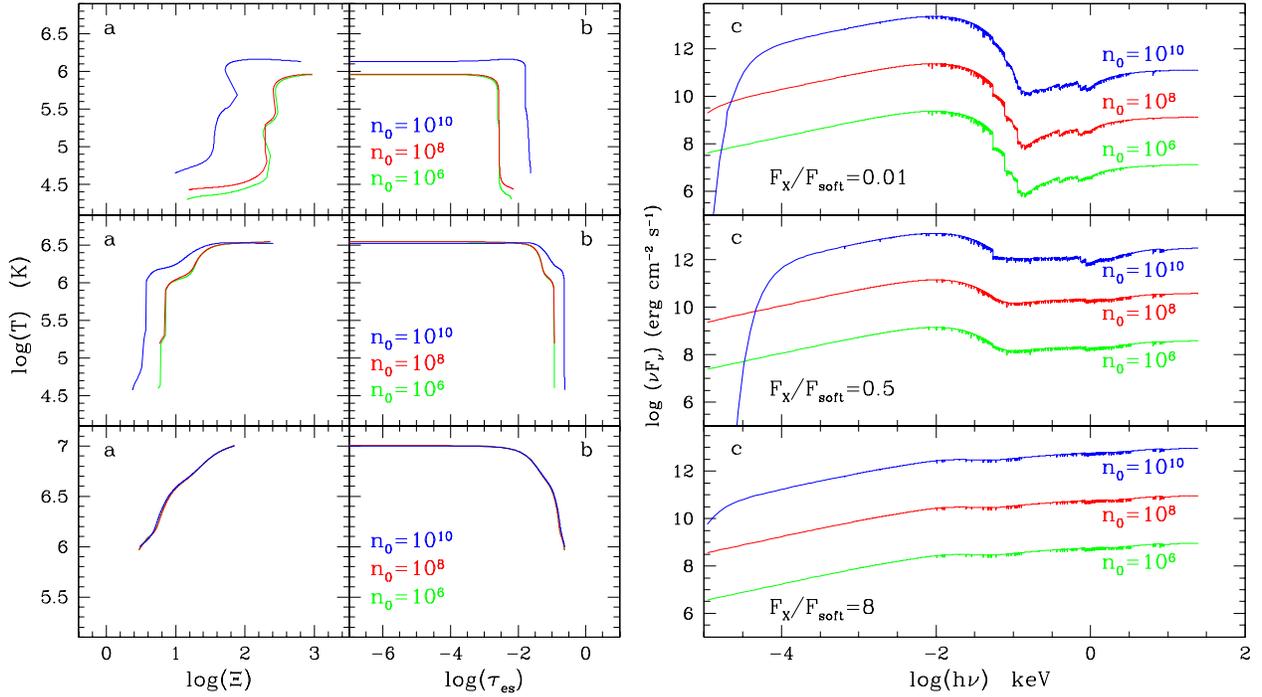}
\caption{ Ionization structures (a), 
temperature structures (b),  and transmitted spectra (c) for
the WA illuminated by double power-law for
three different values of $n_0$: $10^6$~cm$^{-3}$ (green
line), $10^8$~cm$^{-3}$ (red line), and  $10^{10}$~cm$^{-3}$ (blue
line). Upper panels correspond to $F_X/F_{soft}$ equal $0.01$, middle panels
to 0.5, and bottom panels to 8.}
\label{fig:doub}
\end{figure*}

For a single power-law irradiation, 
the primary continuum spectrum ranges from 0.01 up to 100 keV  
and has a spectral index equal to 0.8. 
In the case of a double power-law, the  softer power-law ranges from
$10^{-7}$ up to 0.0136 keV with a spectral index $\alpha_{soft}=0.3$, 
and the hard continuum component has a spectral index $\alpha_X =0.8$, 
ranging from 0.0136 up to 100 keV. 
We consider three cases of relative normalization of 
both power-law components : (i)~$F_X/F_{soft}=0.01$,
corresponding to a domination by the soft component; (ii)~$F_X/F_{soft}=0.5$,
typical of an intermediate case; and (iii)~$F_X/F_{soft}=8$, where
the hard component dominates.
 
\subsection{Hard, single power-law photoionized models}

The comparison of the clouds with different $n_0$ values
 irradiated by a hard, single power-law 
is presented in Fig.~\ref{fig:sing}. Panel~(a) represents the ionization
structure through the WA, which is identical for all $n_0$ values.
The maximum temperature on the illuminated side of the clouds is
quite high, of the order of $4\times 10^7$~K. Going deeper into the 
WA, the temperature structure displays a 
strong gradient (panel (b)), which is consistent with the gas 
structure being in
hydrostatic equilibrium and illuminated
by hard X-rays \citep{rozanska2002,madej2004}.

The transmitted spectra are presented in Fig.~\ref{fig:sing}~(c). 
They differ only in normalization, which means that 
the low density cloud, with $n_0=10^6$~cm$^{-3}$, 
located further away from the continuum source (see Eq.~\ref{eq:ion}),
has the same spectral features as the dense cloud of
$n_0=10^{10}$~cm$^{-3}$, located two orders of magnitude closer to the
source. The detailed comparison of lines for those two cases 
is presented in Fig.~\ref{fig:lin} (bottom panels).   
Lines are identical in all spectral bands,
therefore, models with a single, hard power-law illumination are
highly degenerate and useless for the purpose of determining the
distance to the WA. 

\subsection{Double power-law photoionized models}

The situation changes when we consider a primary continuum described
by a double power-law dominated by its soft component.
In Fig.~\ref{fig:doub} we show results for 
clouds illuminated by a double power-law with increasing 
$F_X/F_{soft}$ ratio, from 0.01 (upper panel) up to 8 (bottom panel). 
When the continuum soft component dominates, the ionization and 
temperature structures of the considered clouds differ for 
different values of $n_0$. Separation between models is clearer for 
lower values of the $F_X/F_{soft}$ ratio. The equilibrium temperature 
at the illuminated face of the cloud is about one order of magnitude
lower than in the case of a single power law illumination. 
 
For the highest value of $n_0=10^{10}$ cm$^{-3}$, the 
continuum spectrum displays a significant absorption in the low energy
band due
to free-free absorption. 
Looking closer at the transmitted spectra, there is a big difference
in the equivalent width (EW) of absorption lines between clouds 
of two extreme values of $n_0$,
and for $F_X/F_{soft}=0.01$; this is illustrated
in Fig.~\ref{fig:lin} upper panels. 
 For instance, EWs of H0$_{\delta}$ and H0$_{\epsilon}$ lines 
for the case of $n_0=10^{10}$ cm$^{-3}$ are ten times smaller than for 
the case of $n_0=10^{6}$ cm$^{-3}$. In X-ray domain, EWs of exampled
lines: 
Fe19$_\delta$, Fe20$_\delta$, Fe17$_{fl}$ and Fe19$_{fl}$ are
respectively seven, five, fourteen and four times larger 
for the cloud with $n_0=10^{10}$ cm$^{-3}$ than for the cloud with 
$n_0=10^{6}$ cm$^{-3}$.
Therefore, from the ratios of EWs of absorption lines
alone, we can select one particular photoionized model which fits the
observations, and then having obtained the number density at the illuminated
cloud surface, we can derive
the distance to the WA.  

The observed differences of photoionized models can be much 
higher for different
spectral energy distributions as shown by \citet{rozanska2007}.
Below, we describe what causes this differences in the case of a
WA. 

\section{Discussion}
\label{sec:dis}

\begin{figure}[t]
\epsfxsize=7cm \epsfbox[260 295 500 690]{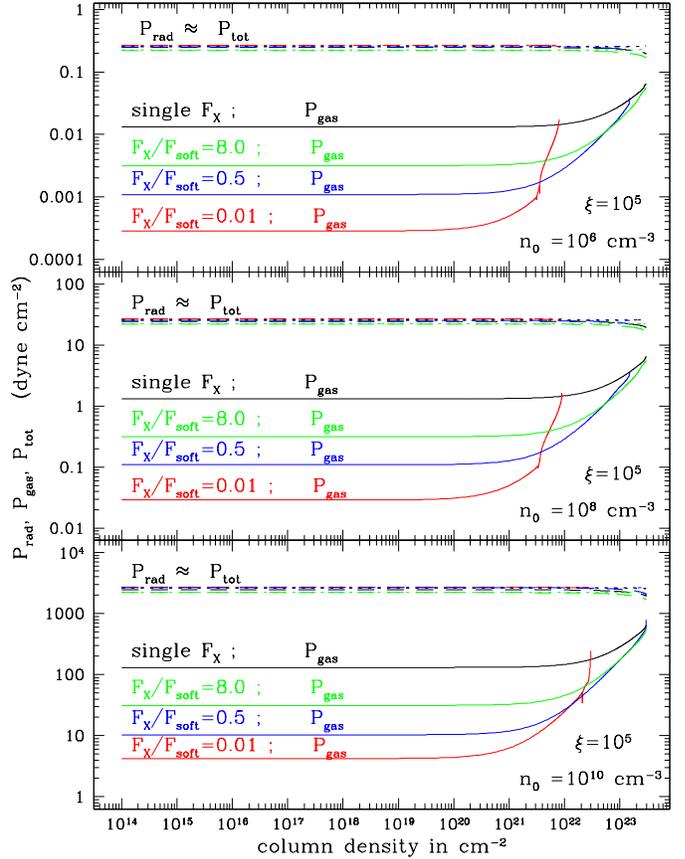}
\caption{Comparison of the gas pressure 
(solid line), the radiation pressure (dashed line), and the total, 
i.e. gas+radiation,  pressure (dotted line), for all computed models. 
The upper panel corresponds to clouds with, 
$n_0=10^6$~cm$^{-3}$, the middle panel to $n_0=10^8$~cm$^{-3}$, 
and the bottom panel to $n_0=10^{10}$~cm$^{-3}$.}
\label{fig:pres}
\end{figure}
\begin{figure}[t]  
\epsfxsize=7cm \epsfbox[260 400 500 700]{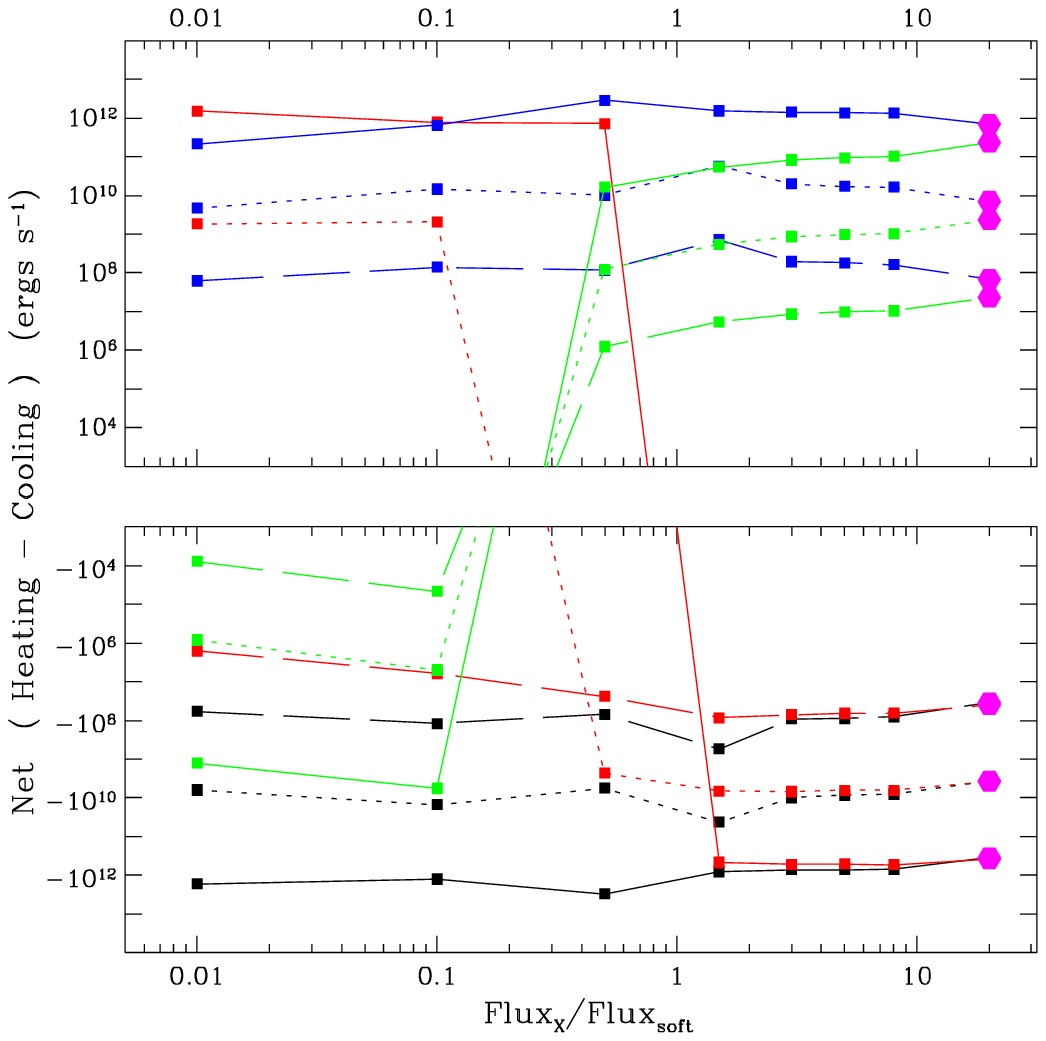}
\caption{Net, i.e. Heating - Cooling gains of energy 
integrated over the cloud versus the ratio of hard X-rays to the soft
power-law component. Net gains for 
Comptonization are represented by green lines, bound-free
processes, i.e. ionization-recombination,  are represented by blue lines, 
bound-bound processes, i.e. line heating-cooling, by black lines, 
and free-free processes by red
lines. Axis are logarithmic, therefore we arbitrarily assigned  
a minus sign to those cases where total gains are negative, i.e. cooling
dominates over heating (bottom panel). There is a 
discontinuity between the cases where heating dominates over 
cooling (upper panel). For each process, solid lines represents clouds 
with $n_0=10^{10}$~cm$^{-3}$, dotted lines $n_0=10^8$~cm$^{-3}$,
and dashed lines  $n_0=10^6$~cm$^{-3}$. 
Magenta hexagons mark gains for the case of a hard, single
power-law illumination.}
\label{fig:bil}
\end{figure}

In order to understand why  single power-law models are degenerate,
and why after including a second illuminating component models can be
effectively separated, 
we conducted an analysis of the  basic physical conditions inside the clouds. 
We followed the behaviour of the parameters responsible for the
hydrostatic and the ionized thermal
equilibrium of the gas, i.e. pressure and efficiency of radiative
processes.   

\subsection{Pressure balance through the cloud}
 
All computed cloud models are dominated by radiation pressure, as
shown in Fig.~\ref{fig:pres}.
For all cases, the radiation pressure (dashed lines) 
practically equals the total (gas + radiation) pressure 
(dotted line).
There is nothing unusual in the behaviour of the gas pressure (solid
lines) for all computed models. For a given $n_0$
the gas pressure is highest for the highest 
$F_X/F_{soft}$. Obviously, the value of the gas pressure on the cloud
surface is proportional to the value of the hydrogen  number density on
the cloud illuminated side. 

\subsection{Energy balance within the cloud}

\begin{figure*}[t]
\epsfxsize=6.6cm \epsfbox[10 330 220 690]{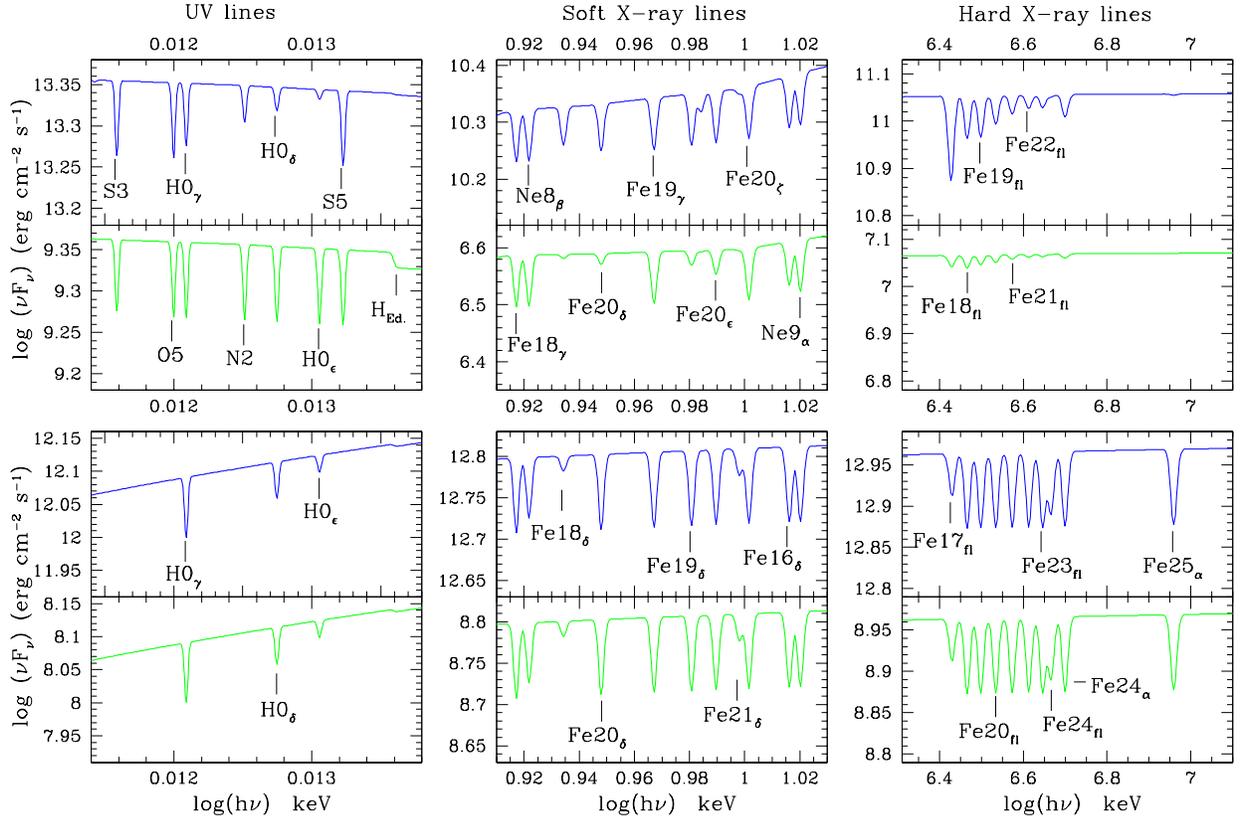}
\caption{Comparison of spectral absorption lines 
for three energy bands: UV (left panels), Soft X-ray 
(middle panels), and Hard X-ray lines (right panels). Upper panels 
show the case where the WA is illuminated by a double power-law
with $F_{X}/F_{soft}=0.01$, while bottom panels correspond to irradiation
by a hard, single power-law. Blue lines depict dense clouds  
of $n_0=10^{10}$~cm$^{-3}$, and green lines less dense clouds  
of $n_0=10^6$~cm$^{-3}$.}
\label{fig:lin}
\end{figure*}

The {\sc titan} code solves the  energy balance by calculating radiation 
gains and losses at each point of the gas slab. 
Fig.~\ref{fig:bil} shows the  
net heating minus cooling rate integrated over the whole cloud versus
the ratio of the two continuum components, $F_X/F_{soft}$.
Each important mechanism was analyzed separately to 
identify its contribution to the total energy
 balance in the WA. 
Net Comptonization is represented in green, ionization minus
recombination in blue, bound-bound heating minus cooling in
black, and finally, free-free balance in red. 

The heating and cooling rates cover a significant range, so we
decided to plot the results on a logarithmic scale.
We arbitrarily assign a minus sign to those cases
where total gains are negative, i.e. when cooling dominates 
(bottom panel). 
The sign is positive when heating dominates (upper panel).  
There is an arbitrarily discontinuity between the two panels in 
Fig.~\ref{fig:bil}, 
which does not have any physical meaning. 
A different type of line is used to denote clouds of different
number densities, $n_0=10^{10}$~cm$^{-3}$ (solid line),
$n_0=10^8$~cm$^{-3}$ (dotted line), 
and $n_0=10^6$~cm$^{-3}$ (dashed line). 
Magenta hexagons represent the 
net heating minus cooling for the case of a single power-law
illumination, they are obvious limits of the radiation rates when 
increasing the $F_X/F_{soft}$ ratio.

Hard X-ray radiation always heats the WA, even if the X-ray
component is very weak. The same is true for the line cooling, which 
most probably is balanced by photoionization. 
The situation changes dramatically for Comptonization and free-free 
processes. For  $F_X/F_{soft}$ decreasing 
from 1 and 0.1, Compton heating
decreases and the WA gas starts to be even Compton cooled by 
soft disk photons. 
The same happens in the case of bremsstrahlung
free-free processes. 
Bremsstrahlung heating starts to dominate when Compton heating becomes
unimportant. 
This means that, in the case of a WA illuminated by a double
power-law, the presence of a soft, disk component changes the mechanism 
responsible for the high temperature equilibrium on the 
irradiated side of the cloud. 
Therefore, in such a situation, clouds are heated mostly by bremsstrahlung, 
and not by Comptonization, like it is in case of clouds illuminated
by a single, hard X-ray continuum.

\subsection{Absorption lines from double power-law models} 

In Fig.~\ref{fig:lin}, we focus on the absorption lines present  
in three spectral bands for clouds with $n_0=10^{10}$~cm$^{-3}$ 
(blue line) and $n_0=10^6$~cm$^{-3}$ (green line). 
The upper panels represent models with $F_X/F_{soft}=0.01$, while 
the bottom panels depict models of clouds illuminated by a single, 
hard X-ray power-law. 

For strongly degenerate models, the absorption line spectra are
identical and there is no
possibility to distinguish them based on the number density $n_0$ 
just by fitting 
photoionization models to the observations.  
The situation is very different in the case of a double power-law illumination. 
The equivalent widths of some absorption 
lines differ for different $n_0$,
giving the possibility of fitting one single photoionization
model to the observed spectrum. Therefore, having a grid of 
photoionized models providing the   
main absorption line ratios, we can estimate the number density at the
cloud illuminated surface, and thus the distance to the WA. 
Indeed, even modeling only the X-ray data of 
any AGN, we have to take into account that the WA is
affected not only by an X-ray
power-law from the corona, but also by a soft component from the  disk. 
This is accounted for in this article and justifies the use of a 
double power-law extending from $10^{-7}$ to 100 keV.

\section{Conclusions}
\label{sec:con}

Photoionized models of a WA usually assume
that the illuminating continuum has the shape of a  single power-law
with the spectral index derived from  X-ray observations of a 
particular AGN. 
In this paper, we have shown that it is important to 
include the broad-band continuum from the active nucleus to achieve a 
proper modeling of 
the transmission spectrum through the ionized WA gas. 

We have computed a set of models assuming illumination of the metal rich
material by a double power-law incident continuum, where one component
mimics the emission from an accretion disk, and the second component
represents the hard X-ray emission coming most probably from a hot corona, or 
magnetic flare. 
This allows to break the degeneracy generally observed in
photoionization models using a single hard X-ray illumination. 
By breaking this degeneracy we observe that
models display different ionization and temperature structures
for clouds with different number densities at the illuminated surface. 
Therefore, transmitted spectra look different and in principle it should be possible 
to identify the photoionized model which better fits a given
observation. 

We have shown that this degeneracy breaks down due to a switch in the radiative 
mechanism heating the ionized cloud when $F_X/F_{soft}$ decreases.
For a single power-law irradiation, hard photons mostly participate to
the Compton heating, and this process becomes responsible for the hot 
equilibrium temperature of the illuminated layers. 
When $F_X/F_{soft}$ decreases, i.e. when the soft component starts to dominate, 
Compton heating is not efficient anymore, since the cloud begins to
be more efficiently  
cooled by Comptonization, and a hot temperature equilibrium is 
established due to bremsstrahlung absorption.

In conclusion, by taking into account broad band illumination from the 
active nucleus it is possible to 
determine the  number density at the illuminated side of WA and thus its
distance from the irradiating continuum source.
This method is complementary to the method based on variability
presented by \citet{netzer2003} and \citet{krongold2005}, 
and can be applied to bright
quasars with a strong disk component. 


\begin{acknowledgements}
We thank Hagai Netzer, Anne-Marie Dumont, and Bo\.zena Czerny
for helpful discussion. This work has been supported by the 
Polish Committee for Scientific Research grant No. 1 P03D 008 29,
by FCT grant  BPD/11641/2002, and has been carried out within
the framework of the  
European Associated Laboratory ``Astrophysics Poland-France''.

\end{acknowledgements}

\bibliographystyle{aa}
\bibliography{myrefsacta}


\end{document}